\documentstyle[11pt,newpasp,twoside]{article}
\def\edcomment#1{\iffalse\marginpar{\raggedright\sl#1\/}\else\relax\fi}
\reversemarginpar

\input psfig

\begin{document}
\title{Radiative Transfer in 3D Numerical Simulations}

\author{Robert Stein}
\affil{Department of Physics and Astronomy, Michigan State University,
East Lansing, MI 48824, USA}
\author{ {\AA}ke Nordlund}
\affil{Niels Bohr Institute for Astronomy, Physics, and Geophysics
Copenhagen, DK}

\begin{abstract}
We simulate convection near the solar surface, where the continuum
optical depth is of order unity.  Hence, to determine the radiative
heating and cooling in the energy conservation equation, we must solve
the radiative transfer equation (instead of using the diffusion or
optically thin cooling approximations).  A method efficient enough to
calculate the radiation for thousands of time steps is needed.  We
assume LTE and a non-gray opacity grouped into 4 bins according to
strength.  We perform a formal solution of the Feautrier equation
along a vertical and four straight, slanted, rays (at four azimuthal
angles which are rotated 15 deg. every time step).   We present details
of our method.   We also give some results: comparing simulated and observed
line profiles for the Sun, showing the importance of 3D transfer for the
structure of the mean atmosphere and the eigenfrequencies of p-modes,
illustrating Stokes profiles for micropores, and analyzing the effect
of radiation on p-mode asymmetries.
\end{abstract}

\section{Introduction}

Radiative energy exchange is critical in determining the structure of
the upper convection zone.  Near the surface of the Sun, the energy
flow changes from almost exclusively convective below the surface to
radiative above the surface.  The interaction between convection and
radiation near the surface determines what we observe, drives the
convection, and generates the magnetic and non-magnetic activity which
heats the chromosphere and corona.  Hence, the interaction between
convection and radiation has significant impact on both the dynamics of
convection and our diagnostics.  Escaping radiation cools the plasma
that reaches the surface, which produces the low entropy plasma whose
buoyancy work drives the convection.  Radiation transport determines
(with convection and waves) the mean atmospheric structure.  Radiation
provides the diagnostics that we use to determine the velocity,
density, temperature and magnetic field on the Sun.  Radiation
transport modifies the observable p-mode temperature fluctuations so as
to reverse the asymmetry of the intensity vs. velocity spectral peaks.

Since the top of the convection zone occurs near the level where the
continuum optical depth is one, neither the optically thin nor the
diffusion approximations give reasonable results.  We need to solve the
3D, LTE, non-gray radiation transfer in our models (Nordlund 1982).

\section{Radiative Heating/Cooling}

Radiation enters the calculation through the radiative heating/cooling
term in the energy equation,
\begin{equation}
{{\partial e}\over{\partial t}} + {\mathbf v}\cdot\nabla e =
-{{P}\over{\rho}} \nabla\cdot P + Q_{\rm rad} + Q_{\rm dissipation}
\ ,
\end{equation}
where the radiative heating/cooling is given by
\begin{equation}
Q_{\rm rad} = 4 \pi \int_{\lambda} \kappa_{\lambda}
\left(J_{\lambda}-S_{\lambda}\right) d \lambda
\ .
\end{equation}
We calculate $J_{\lambda}-S_{\lambda}$ by a formal solution of the
Feautrier equations assuming LTE so the source function is the Planck
function,
\begin{displaymath}
{{d^2 P_{\lambda}}\over{d \tau_{\lambda}^2}} = P_{\lambda}-B_{\lambda}
\end{displaymath}
where $P_{\lambda}$ is the Feautrier mean intensity
\begin{displaymath}
P_{\lambda} =
   {1\over{2}}\left[I_{\lambda}(\Omega)+I_{\lambda}(-\Omega)\right]
   \ .
\end{displaymath}

\section{Opacities}
%
Continuous opacity is calculated explicitly using the Uppsala package
(Gustafsson et al. 1975), for the entire temperature-density range in the
simulation and stored in a table.  To account for UV
opacities not in the original package, the UV continuous opacity is
enhanced below 500 nm by a function of $\lambda$, according to
the prescription of Magain (1983).  Line opacities for the
mean 1D atmosphere state are calculated using the Uppsala ODF tables
(Gustafsson et al. 1975).  For a grid of temperatures of the mean atmosphere
state, the Rosseland mean with and without lines is calculated.  The
ratio of full to continuous Rosseland opacities is then used to enhance
the continuous Rosseland opacity at each density in the table for the
given temperature.  In addition, for small optical depths, the
continuum Rosseland mean opacity is corrected by the ratio of the
intensity weighted opacity for the 1D plane parallel average atmosphere
to the continuum Rosseland opacity.  Thus the opacity stored in the table
is
\begin{equation}
\kappa = x_{\kappa}\kappa_{\rm cR} \ ,
\end{equation}
where the correction factor $x_{\kappa}(T)$ is
\begin{equation}
x_{\kappa} = e^{-30\tau_{\rm cR}}
{{\left<\kappa\right>_{J}}\over{\left<\kappa\right>_{\rm cR}}} +
\left(1-e^{-30\tau_{\rm cR}}\right)
{{\left<\kappa\right>_{\rm (L+c)R}}\over{\left<\kappa\right>_{\rm cR}}}
\ .
\label{kappaeff}
\end{equation}
The continuum Rosseland opacity is
\begin{displaymath}
    {1\over{\left<\kappa\right>_{\rm cR}}} =
          {{{\displaystyle \sum_{j(i=1)}}
             {{1}\over{\kappa_{\lambda_j} + \sigma_{\lambda_j}}}
             {{\partial B_{\lambda_j}}\over{\partial T}}w_{\lambda_j}}
    \over
           {{\displaystyle \sum_{j(i=1)}}
           {{\partial B_{\lambda_j}}\over{\partial T}}w_{\lambda_j}}} \ ,
\end{displaymath}
where $j(i=1)$ is the set of all those wavelengths belonging to the
continuum.  The intensity mean is
\begin{displaymath}
    \left<\kappa\right>_{J}=
      {{{\displaystyle\sum_j}\kappa_{\lambda_j} J_{\lambda_j}
                        e^{-\tau_{\lambda_j}/2}w_{\lambda_j}}\over
       {{\displaystyle\sum_j}J_{\lambda_j}
                        e^{-\tau_{\lambda_j}/2}w_{\lambda_j}}} \ .
\end{displaymath}

To summarize:
the opacity table is calculated from the Rosseland mean opacity,
based on the continuum opacity alone, $\left<\kappa\right>_{\rm cR}$.
This is then corrected by the ratio $x_\kappa$ (Eqn~\ref{kappaeff}),
to account for lines and the transition to optically thin radiation,
but this ratio is evaluated only for the $\rho$/$T$-track of the mean
stratification of the simulation (not for the whole table).  The
correction factor, $x_\kappa$, is then regarded as a function of
temperature alone and used to correct the continuum Rosseland opacities
for all densities in the table at each table temperature.

\begin{figure}[!hbt]
\centerline{
\psfig{figure=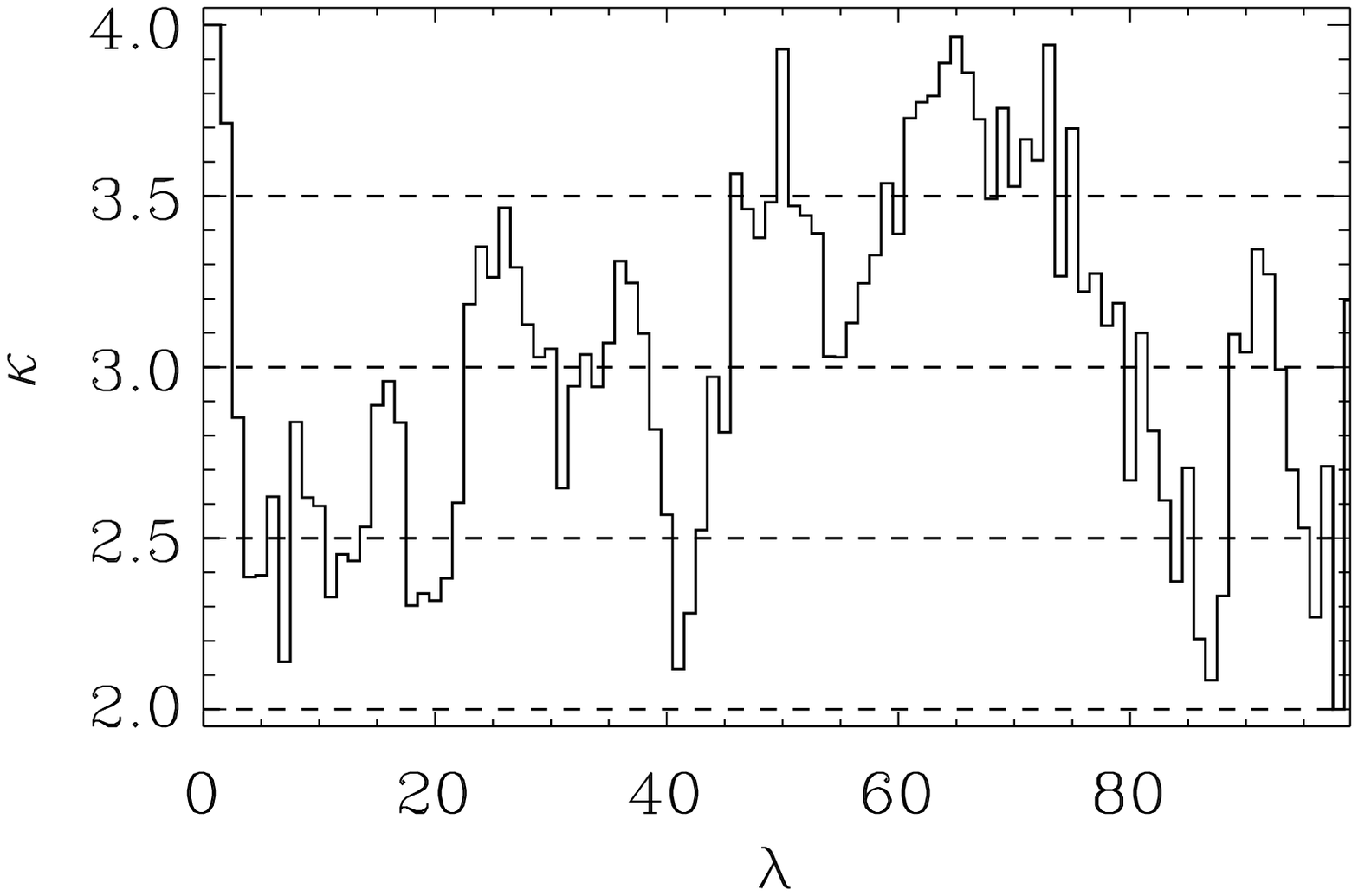,width=6.7cm}
\quad
\psfig{figure=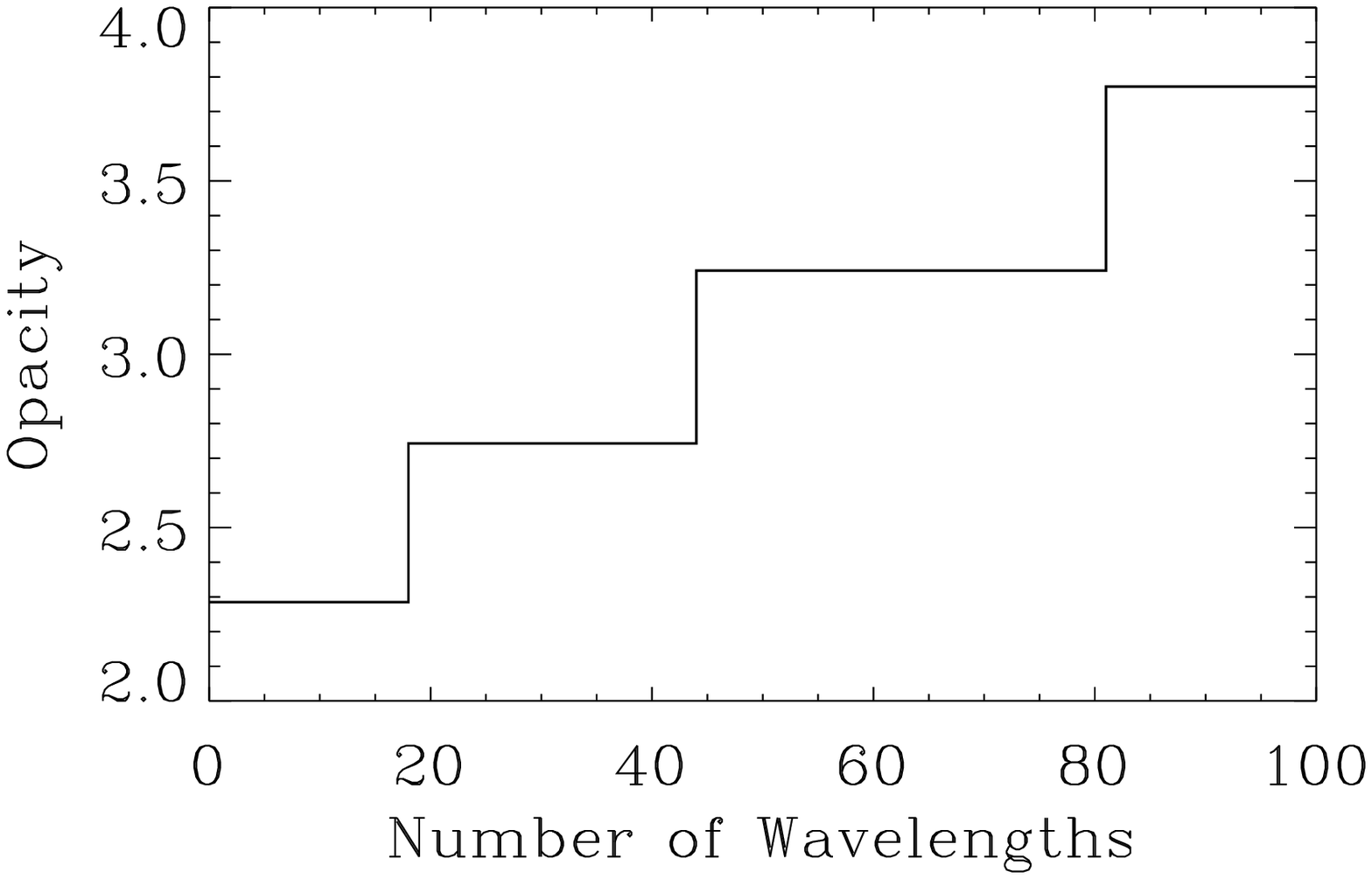,width=6.7cm}
}
\caption{Opacity as a function of wavelength (left) and grouped into
four bins (right).  Each bin has the average value of the opacity in
that bin and the weight equal to the sum of the weights of the
wavelengths belonging to that bin.}
\label{opac}
\end{figure}
The radiation calculation is greatly sped up by drastically reducing
the number of wavelengths at which we solve the Feautrier equations.
We make the simplifying approximation of grouping the opacity at each
wavelength into 4 bins according to its magnitude
(Nordlund 1982, Skartlien 2000).  Suppose the opacity
is as shown in the toy example (Fig.~\ref{opac}).  Each wavelength
is assigned a bin according to the magnitude of its opacity.  In this
toy example, those wavelengths with opacity between 2.0 and 2.5 are
assigned to the first opacity bin with average opacity 2.3 and
considered to represent the continuum (Fig~\ref{opac}).  Those
wavelengths with opacity between 2.5 and 3.0 are assigned to bin two
with average opacity 2.7 corresponding to weak lines.  Those
wavelengths with opacity between 3.0 and 3.5 are assigned to bin three
with average opacity 3.2 corresponding to intermediate strength lines.
Those wavelengths with opacity greater than 3.5 are assigned to bin 4
with average opacity 3.8 corresponding to strong lines.
In practice, we take the opacity bins to have a ratio of a factor of 10
in magnitude (Nordlund 1982),
\begin{displaymath}
\kappa_i = 10^{i-1} \kappa_1 \ .
\end{displaymath}
The weight of each opacity bin is the sum of the weights of each
wavelength in that bin,
\begin{displaymath}
w_i = \Sigma_{j(i)} w_{\lambda_j} \ ,
\end{displaymath}
where $j(i)$ is the non-contiguous set of those wavelengths $\lambda_j$
in bin $i$.  The consequence of this grouping of opacities is that all
wavelengths in a given bin have optical depth unity at approximately
the same geometrical depth so that integrals over optical depth commute
with the sum over wavelengths.  Specifically, the radiation
heating/cooling is
\begin{eqnarray*}
Q_{\rm rad} & = & \int_{\lambda}
  \kappa_{\lambda}\left(J_{\lambda}-B_{\lambda}\right) d \lambda\\
& = & \sum_{i} \sum_{j(i)} \kappa _{\lambda_j}
  \left(J_{\lambda_j} - B_{\lambda_j}\right) w_{\lambda_j}\\
& = & \sum_{i} \sum_{j(i)} \kappa_{\lambda_j} L_{\tau_{\lambda_j}}
  \left(B_{\lambda_j}\right)w_{\lambda_j}\\
& \approx & \sum_i L_{\tau_i}
  \left(\sum_{j(i)}B_{\lambda_j}w_{\lambda_j}\right)\\
& \equiv & \sum_i L_{\tau_i} (B_i)w_i\\
& \equiv & \sum_i \kappa_i \left(J_i - B_i\right) w_i \ ,
\end{eqnarray*}
where the operator $L_{\tau_{\lambda}}$ is
\begin{eqnarray*}
J_{\lambda} - B_{\lambda} =
L_{\tau_{\lambda}}(B_{\lambda})
  = \int_0^1 {{d\mu}\over{\mu}}e^{\tau_{\lambda}/\mu}
    \int_0^{\infty} dt e^{-t/{\mu}}B_{\lambda}(t) - B_{\lambda} \ ,
\end{eqnarray*}
and $B_i$ is the weighted sum of the Planck functions at the
wavelengths in bin $i$,
\begin{eqnarray*}
B_i = \left(\sum_{j(i)} B_{\lambda_j} w_{\lambda_j}\right)/w_i \ .
\end{eqnarray*}
This results in a tremendous saving in computational time, since the
Feautrier equations need only be solved for a few opacity groups.  We
choose to use only 4 groups corresponding to the continuum, weak,
medium and strong lines.

A major approximation we make is to assume that bin membership of each
wavelength is independent of temperature.  This is reasonable to the
extent that the ratio of line to continuum opacity is relatively
independent of the temperature, which holds for lines of atomic
species that are also major electron contributors to the dominant
continuous H$^{-}$ opacity.  However, it is not a good approximation
for the numerous iron lines that produce significant line blocking in
the photosphere.

\section{Solution of Feautrier Equations}
\begin{figure}
\centerline{\psfig{figure=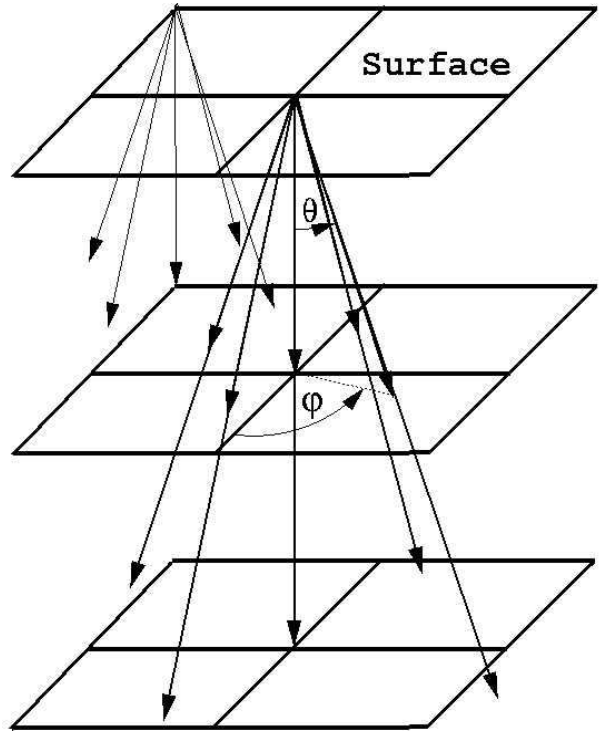,width=8.0cm}}
\caption{The Feautrier equation is solved along one vertical and 4
inclined rays through each point on the surface.  Values of the opacity
and source function are interpolated to the intersection of the
ray with each horizontal plane.  The azimuthal angle $\phi$ of the
inclined rays is rotated 15 degrees each time step in order to sample
the surrounding medium.}
\label{rays5}
\end{figure}

We solve the Feautrier equations along 5 rays -- one vertical ray and
four slanted rays (at an angle $\cos \theta = \mu = 1/3$), through each
grid point on the surface, using long characteristics (Fig.~\ref{rays5}).
The vertical
ray has weight 1/4 and the slanted rays together weight 3/4.  The four
slanted rays are rotated by 15 degrees each time step in order
effectively probe all the surrounding medium.
Each $\Theta, \phi$ combination is treated in turn.  For each, there is
a collection of parallel rays labeled by the surface grid point through
which they pass.  For the slanted rays, the opacity and source function
are interpolated to the location of the intersections of the rays with
each horizontal plane.  The optical depth along each ray at each
horizontal plane is then calculated by integrating along the ray from
the surface inward.  Since the opacity groups are fixed multiples (10)
of each other, the optical depth needs to be calculated for the
continuum only and is then scaled for the other opacity groups.

To obtain accurate integrated intensities the continuum opacity average
is renormalized slightly, with a factor (close to unity) obtained by
requiring that the emergent intensity, summed over the four bins, should
match the wavelength integrated monochromatic intensity, computed in a
two-dimensional strip from the model.  A posteriori checks, where similar
comparisons are made for 3-D snapshots, have been used to verify the
procedure.

It is crucial that the thin surface thermal boundary layer be resolved
in order for the correct radiative losses to be obtained.  Therefore,
in low resolution simulations we interpolate from the hydrodynamic
computational grid to a finer grid in the vertical direction for
solving the radiation transport.  In high resolution simulations this
is not necessary because the thermal boundary layer is already resolved
on the dynamic grid.

At large optical depths, $P_{\lambda} \rightarrow B_{\lambda}$, which
leads to roundoff errors in the heating $q_{\lambda} = P_{\lambda}-B_{\lambda}$.
We therefore solve the Feautrier equation directly for
$q_{\lambda} = P_{\lambda}-B_{\lambda}$, by rewriting it as
\begin{equation}
{{d^2 q_{\lambda}}\over{d \tau_{\lambda}^2}} = q_{\lambda} -
{{d^2 B_{\lambda}}\over{d \tau_{\lambda}^2}}
\ .
\end{equation}
This ensures that the radiative heating/cooling goes to the correct
asymptotic solution at large optical depth.  The boundary conditions are
\begin{eqnarray*}
{{d q_{\lambda}}\over{d \tau_{\lambda}}} & = & q_{\lambda} + B_{\lambda}
- {{d B_{\lambda}}\over{d \tau_{\lambda}}}
\qquad (\tau\rightarrow 0)\\
q_{\lambda} & \rightarrow 0, & \qquad (\tau\rightarrow\infty)
\end{eqnarray*}

The finite difference form of the Feautrier equation is,
\begin{eqnarray*}
\lefteqn{
q_{j-1}\left({{1}\over{\tau_{j}-\tau_{j-1}}}\right.
  \left.{{2}\over{\tau_{j+1}-\tau_{j-1}}}\right)
 -q_{j}\left[1+\left({{1}\over{\tau_{j}-\tau_{j-1}}}+
  {{1}\over{\tau_{j+1}-\tau_{j}}}\right)
  {{2}\over{\tau_{j+1}-\tau_{j-1}}}\right]}\\
& & +q_{j+1}\left({{1}\over{\tau_{j+1}-\tau_{j}}}
   {{2}\over{\tau_{j+1}-\tau_{j-1}}}\right) = \\
& & S_{j-1}\left({{1}\over{\tau_{j}-\tau_{j-1}}}\right.
  \left.{{2}\over{\tau_{j+1}-\tau_{j-1}}}\right)
  -S_{j}\left[\left({{1}\over{\tau_{j}-\tau_{j-1}}}+
  {{1}\over{\tau_{j+1}-\tau_{j}}}\right)
  {{2}\over{\tau_{j+1}-\tau_{j-1}}}\right]\\
& & S_{j+1}\left({{1}\over{\tau_{j+1}-\tau_{j}}}\right.
  \left.{{2}\over{\tau_{j+1}-\tau_{j-1}}}\right)
  \ .
\end{eqnarray*}
This has the tri-diagonal form
\begin{displaymath}
A_{j}q_{j-1}+B_{j}q_{j}+C_{j}q_{j+1}= D_{j}
\ .
\end{displaymath}

There is a roundoff error problem at small optical depth, because
the $1$ in $B_{j}$ becomes so small compared to $1/\Delta \tau^{2}$
that it is lost in the noise and the solution becomes indeterminate.
As a result, this formulation can only be used down to optical depths
the order of the square root of the machine word length.
The solution is to store the sum of the tridiagonal matrix
elements, rather than $B_{j}$, in the diagonal slot
(Nordlund 1982, Rybicki \& Hummer 1991),
initializing it with
\begin{displaymath}
D_{j} == A_{j}+B_{j}+C_{j} = -1
\end{displaymath}
The tridiagonal system is solved by the standard elimination
procedure, except that the diagonal element $B_{j}$ is recovered
just before the back-substitution step, where it may be obtained
from
\begin{displaymath}
B_{j} = D_{j}-C_{j}
\ ,
\end{displaymath}
since at that point $A_{j} = 0$ from the forward elimination.

With these substitutions, the solution is accurate for arbitrarily
small optical depths.

Total radiative heating/cooling at each location where a ray crosses a
horizontal coordinate surface is accumulated as the sum over opacity bins
($i=1,4$), and angles of the Feautrier $q_{i}(\theta, \phi)$,
\begin{equation}
Q_{\rm rad} = 4 \pi \sum_{\Omega} \sum_{i} \kappa_{i}
q_{i}(\Omega) w_{i} w_{\Omega} \ .
\end{equation}
The radiative heating/cooling is finally interpolated from the
slanted grid back to the Cartesian computational grid and added into
the expression for the time derivative of the internal energy and the
variables are advanced in time.

\section{Interaction of Radiation and Convection}

The interaction between radiative and convective energy transport has
a profound effect on the structure, dynamics and diagnostics of the Sun.
Escaping radiation cools the plasma
that reaches the solar surface and produces the low entropy plasma whose
buoyancy work drives the convection.  Radiative, convective and wave
energy transport determines the mean structure of the solar atmosphere.
Radiation emerging  from the atmosphere
provides the diagnostics that we use to determine the velocity,
density, temperature and magnetic field on the Sun.  Radiation
transport modifies the observable p-mode temperature fluctuations so as
to reverse the asymmetry of the intensity vs. velocity spectral peaks.

\subsection{Convection Driver}

\begin{figure}[!htb]
\centerline{\psfig{figure=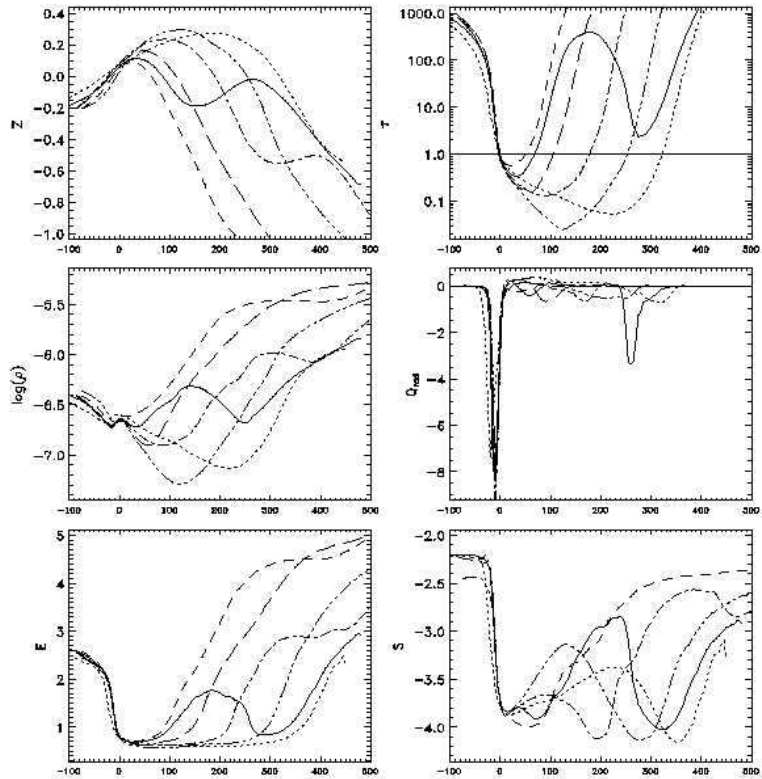,width=10.0cm}}
\caption{Fluid that approaches the surface is cooled by radiating away
its thermal and ionization energy.  This decrease its entropy and it is
pulled down by gravity to form the cores of downflow plumes.  The
largest entropy fluctuations and most of the buoyancy work that drives
the convection occurs in the downflows.
}
\label{parcel_history}
\end{figure}
Solar convection is driven primarily from a thin surface thermal
boundary layer by radiative cooling.  Fluid that approaches the surface
loses its thermal and ionization energy to escaping radiation
(Fig~\ref{parcel_history}).  This reduces its entropy, so gravity pulls
it down to form the cores of the downflows.  Large entropy fluctuations
occur only in these downflows, so that most of the buoyancy work that
drives the convection occurs in these downflows.  Thus the primary
driver of solar convection is radiative cooling from the surface
(Stein \& Nordlund 1998).

\subsection{Atmospheric Structure}

\begin{figure}[!htb]
\centerline{\psfig{figure=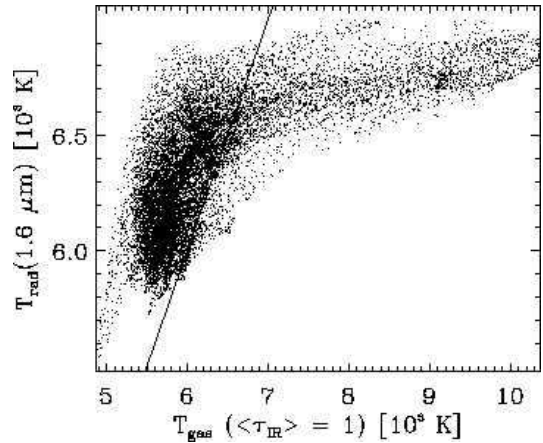,width=7.0cm}}
\caption{Radiation temperature in the IR at 1.6 ${\mu}$m vs. gas
temperature at $<\tau_{IR}> = 1$.  Hot gas is not observed.  The large
temperature sensitivity of the H$^-$ opacity leads to large optical
depths where the gas is hot and hides it from view.  Thus the average
gas temperature is higher than the observed radiation temperature, so
the 3D atmosphere is more extended than a 1D atmosphere with the same
effective temperature.}
\label{trad_tt}
\end{figure}
\begin{figure}[!htb]
\centerline{\psfig{figure=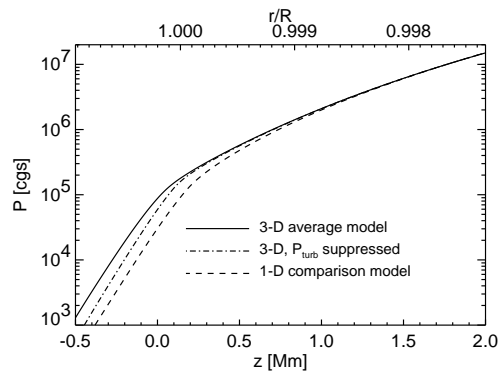,width=7.0cm}}
\caption{Atmospheric stratification is extended $\sim$ 150 km, half by
3D radiative transfer effects (Fig~\ref{trad_tt}) and half by turbulent
pressure support.}
\label{elev-P}
\end{figure}

The average structure of the solar atmosphere is significantly
different from that of 1D models that reproduce the observed solar
luminosity and radius.  First, due to 3D radiative transfer effects the
average temperature of a 3D model is higher than a 1D model with the
same effective temperature.  The great temperature sensitivity of the
H$^-$ opacity makes the optical depth of hot gas very large, so it is
not observed (Fig~\ref{trad_tt}).  Hence, the actual average gas
temperature is higher than the observed radiation temperature.  A hotter
atmosphere has a larger pressure scale height and is more extended.  In
the solar case this gives an extra 75 km extension to the atmosphere.
Second, turbulent pressure gives extra support to the atmosphere, which
extends it an additional 75 km. (Fig~\ref{elev-P})
(Rosenthal et al. 1999).

\subsection{Atmospheric Diagnostics}

\begin{figure}[!htb]
\centerline{
\psfig{figure=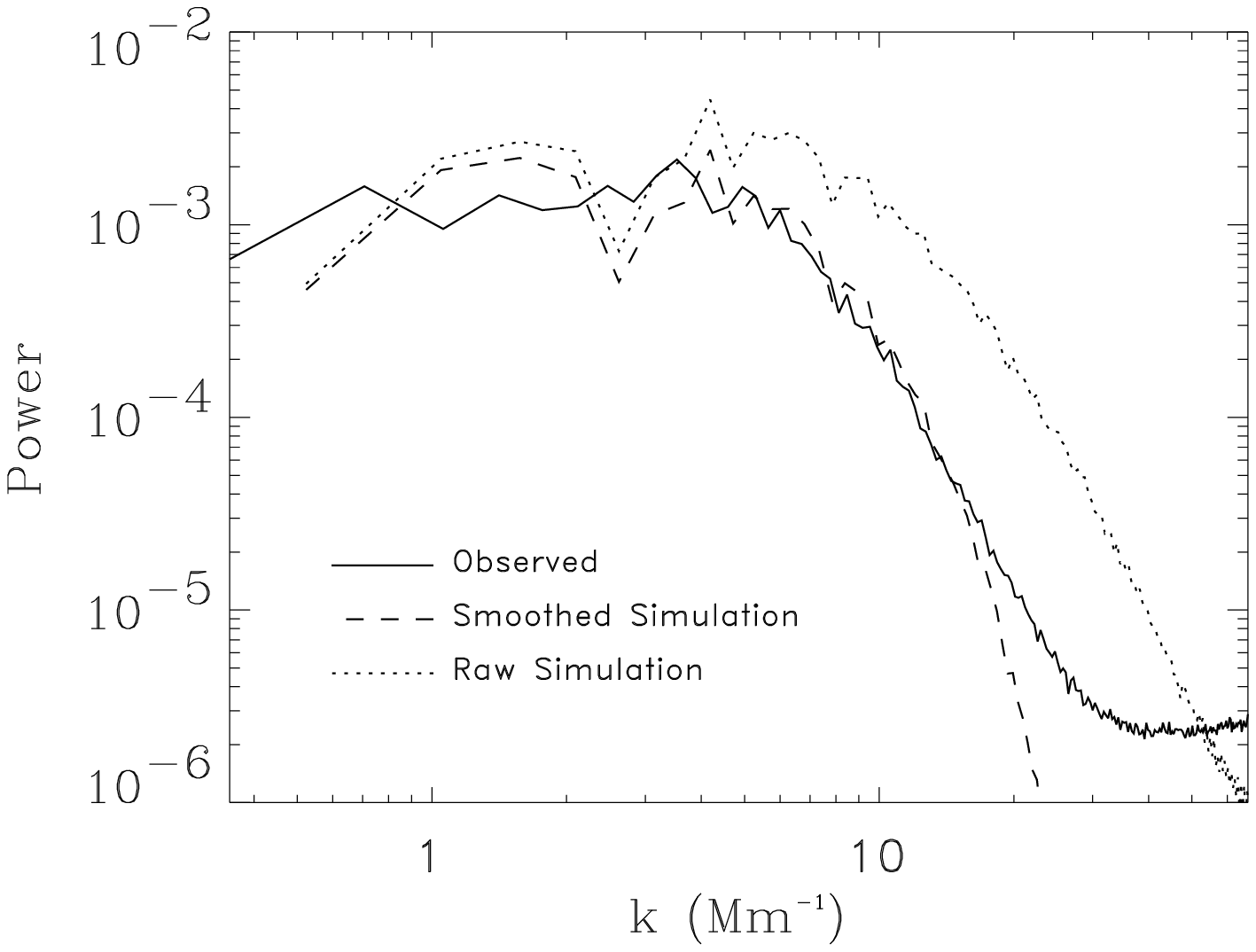,width=6.7cm}
\quad
\psfig{figure=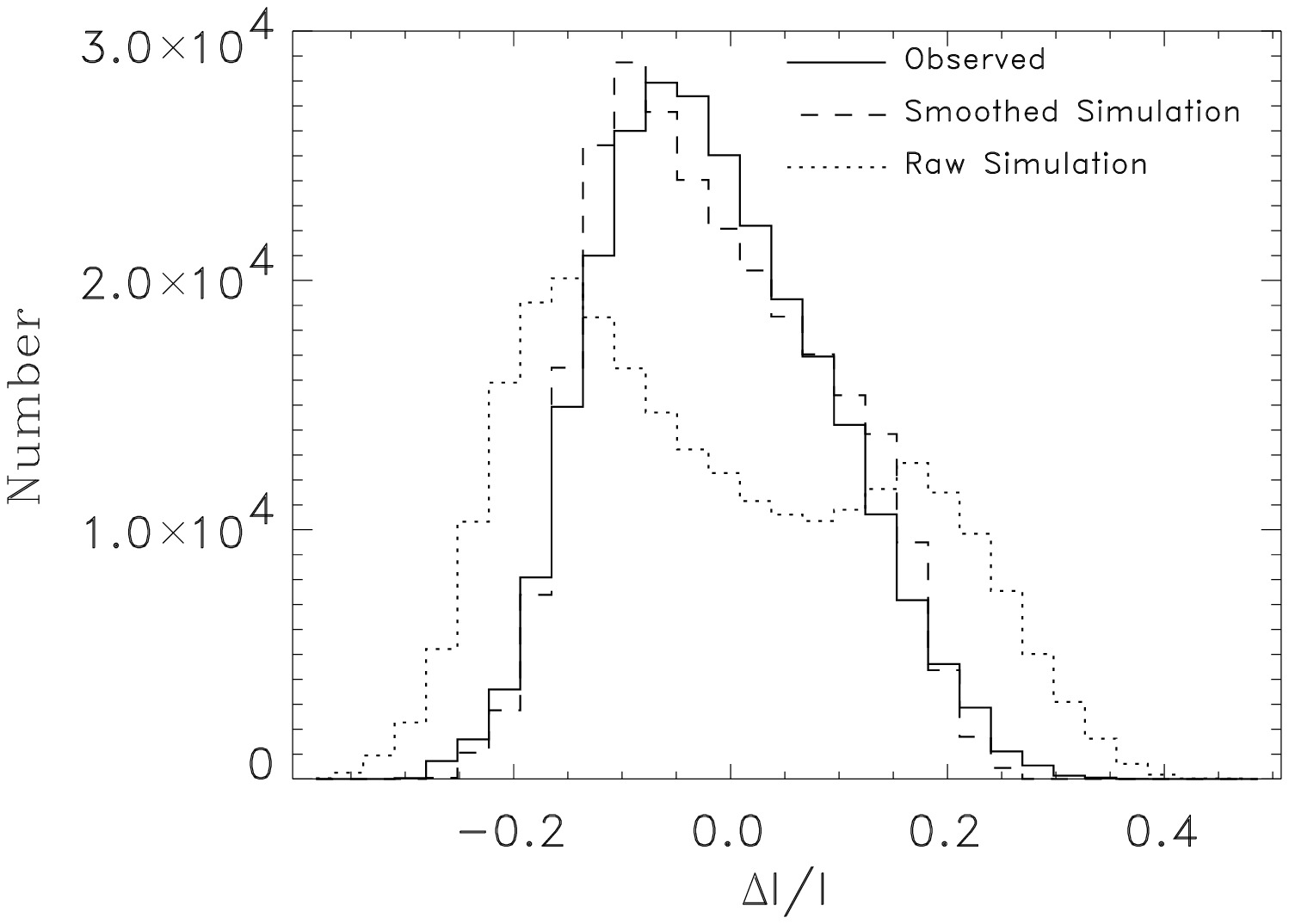,width=6.7cm}
}
\caption{Granulation size spectrum and emergent intensity distribution
compared to observed values.}
\label{granulation}
\end{figure}

Emerging continuum and line radiation provides us with the diagnostic
tools to determine the properties of the solar atmosphere.  Emergent
radiation from the solar surface can be resolved spatially and
temporally.  In the continuum radiation we see the granulation.  The
spatial spectrum (Fig~\ref{granulation}) gives the power in the different
scales of motion.  The intensity distribution tells us about the
temperature contrast on the unit optical depth surface.

\begin{figure}[!htb]
\centerline{
\psfig{figure=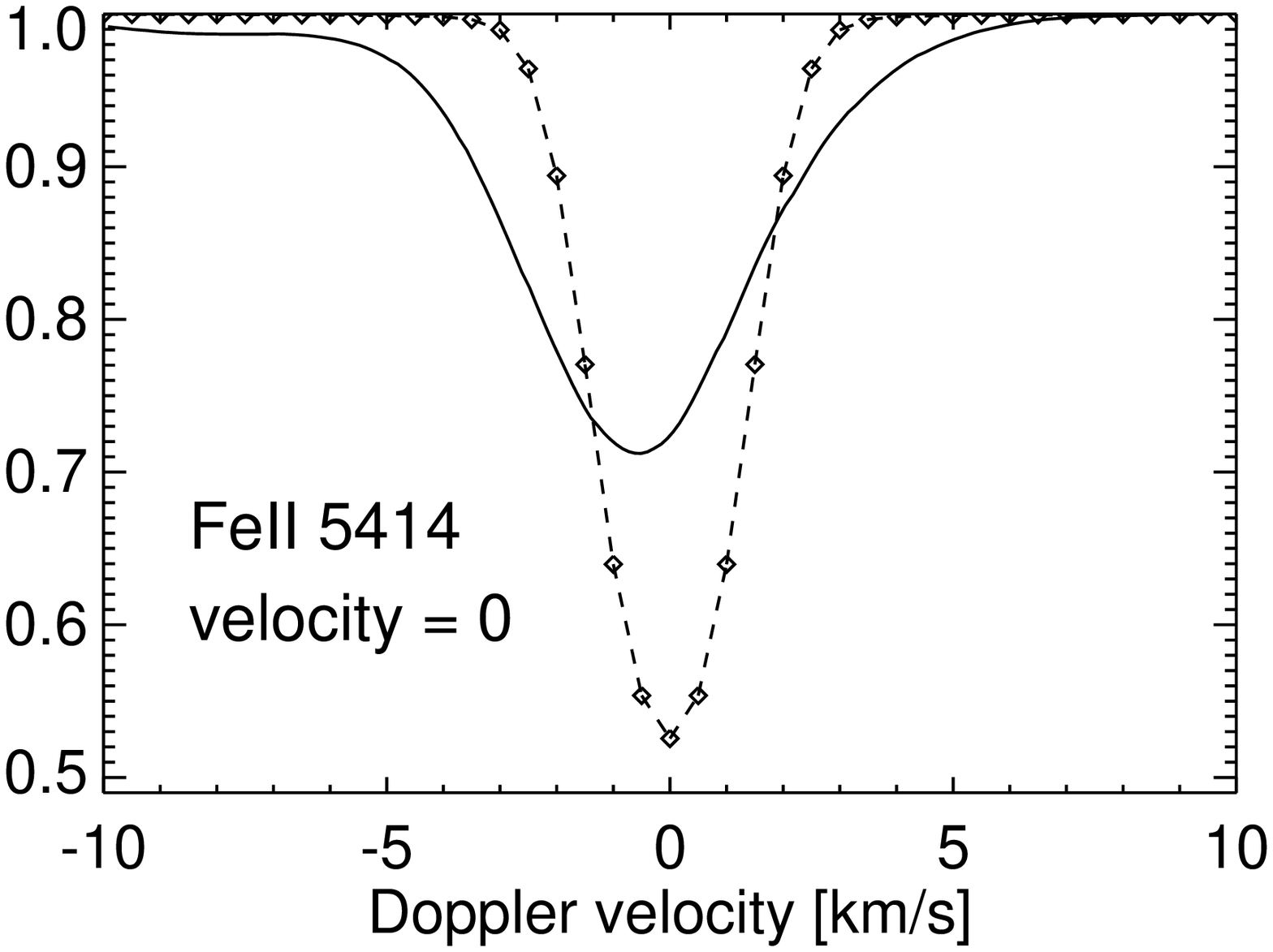,width=6.3cm}
\qquad
\psfig{figure=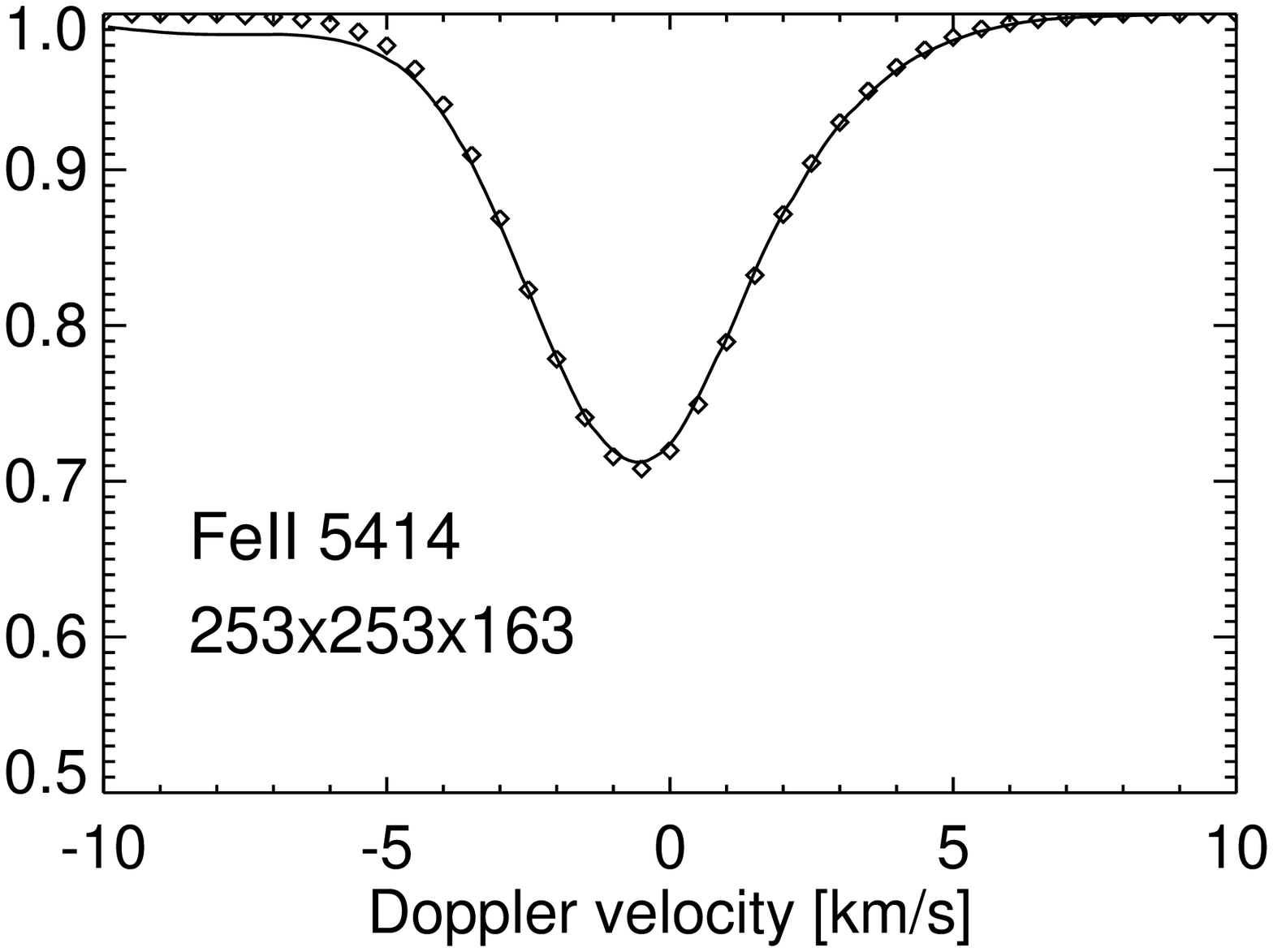,width=6.3cm}}
\caption{Average line profile for Fe $\lambda$ 541.4 line, with
temperature structure from the simulation, with no velocities (left) and
with the full velocity field (right).}
\label{line_profiles1}
\end{figure}

\begin{figure}[!htb]
\centerline{\psfig{figure=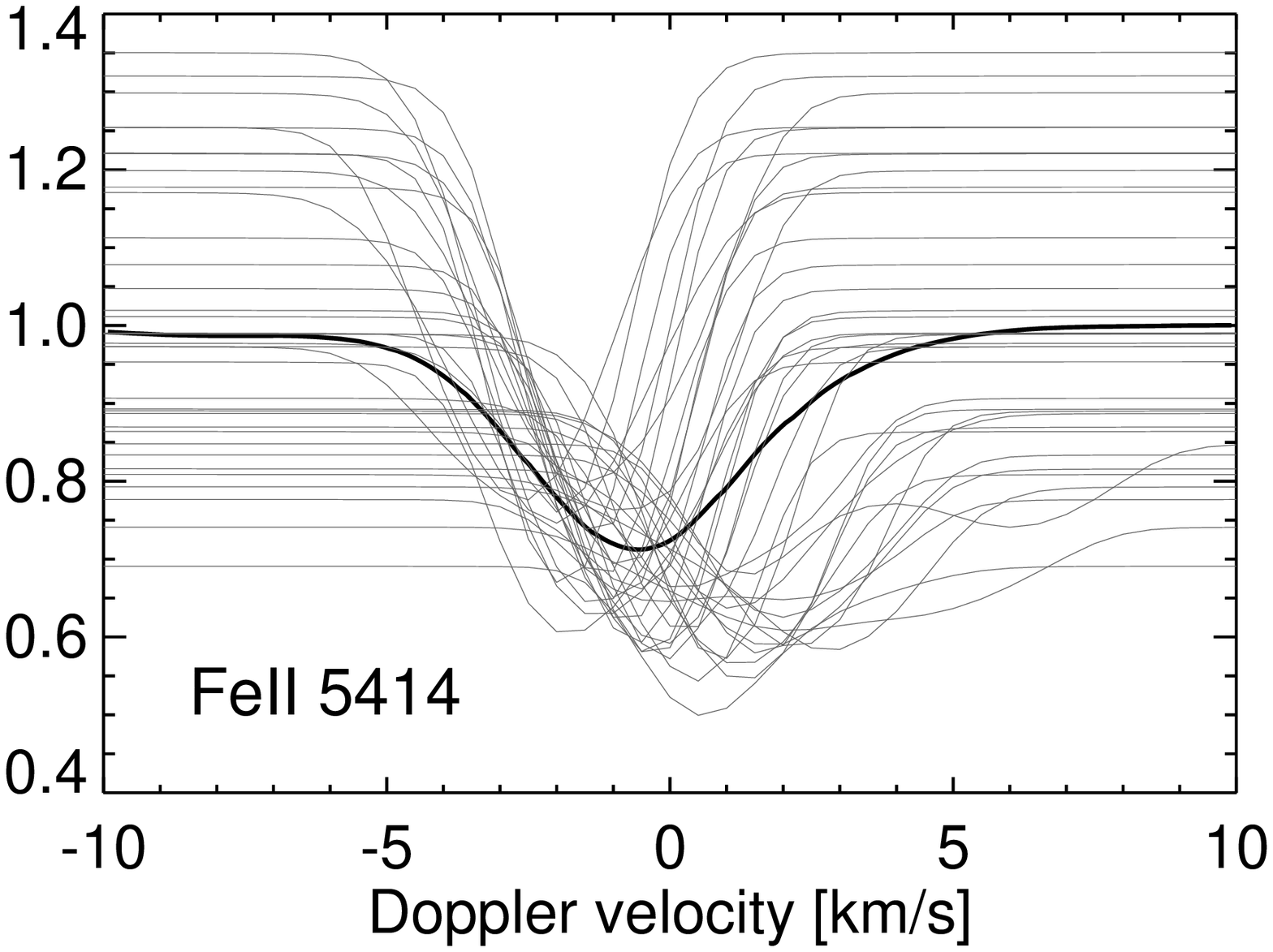,width=7.7cm}}
\caption{The average line profile (heavy line and
Fig~\ref{line_profiles1}) is the combination of profiles from different
locations with different line shifts and widths.  The effects that are
modeled by micro- and macro- turbulence in 1D are all due to the
overshooting convection velocities.}
\label{line_profiles2}
\end{figure}

Line radiation gives us much more information.  Spectral lines of heavy
elements, whose thermal Doppler widths are small compared to typical
photospheric velocities, provide direct diagnostics of velocity and
temperature fluctuations in the photosphere.  Non-spatially resolved
properties such as average line profiles are useful because they bypass
the difficulties associated with atmospheric seeing and instrumental
resolution.  Their shape critically depends on the solar velocity field
-- the temperature structure alone gives lines that are too narrow and
deep (Fig~\ref{line_profiles1} (left)).  Including the convective
overshoot velocities in the photosphere gives excellent agreement with
the observed profiles (Fig~\ref{line_profiles1} (right)).  This
observed average profile is the result of spatial and temporal
averaging of lines from different locations with different temperatures
and line of sight velocities  and very different shifts, widths and
shapes (Fig~\ref{line_profiles2}).  The average line profile results
entirely from convection induced temperature and velocity fluctuations,
without any need for micro- or macro- turbulence or extra damping
(Stein \& Nordlund 2000).

\begin{figure}[!htb]
\centerline{
\psfig{figure=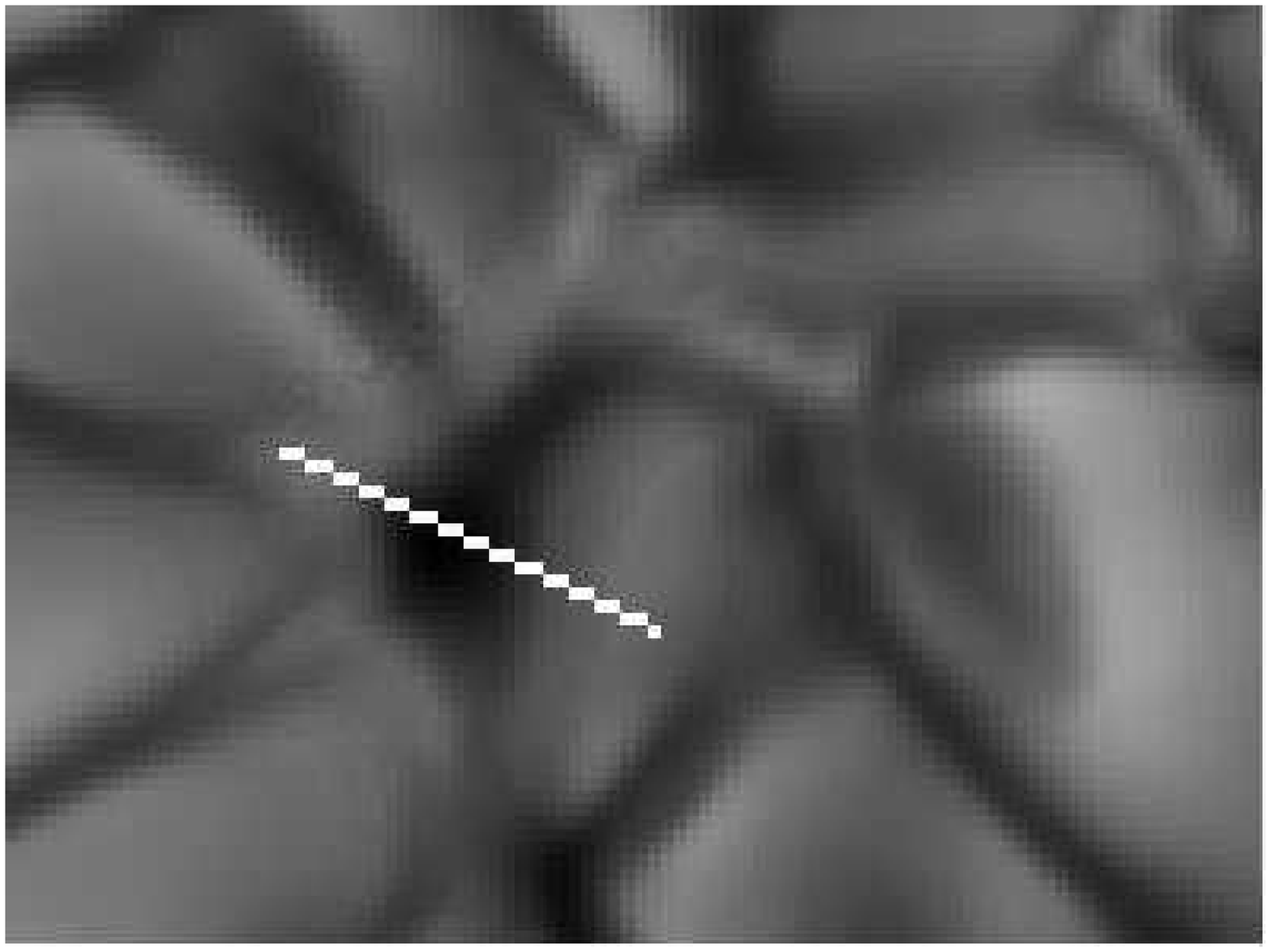,width=4.5cm}
\quad
\psfig{figure=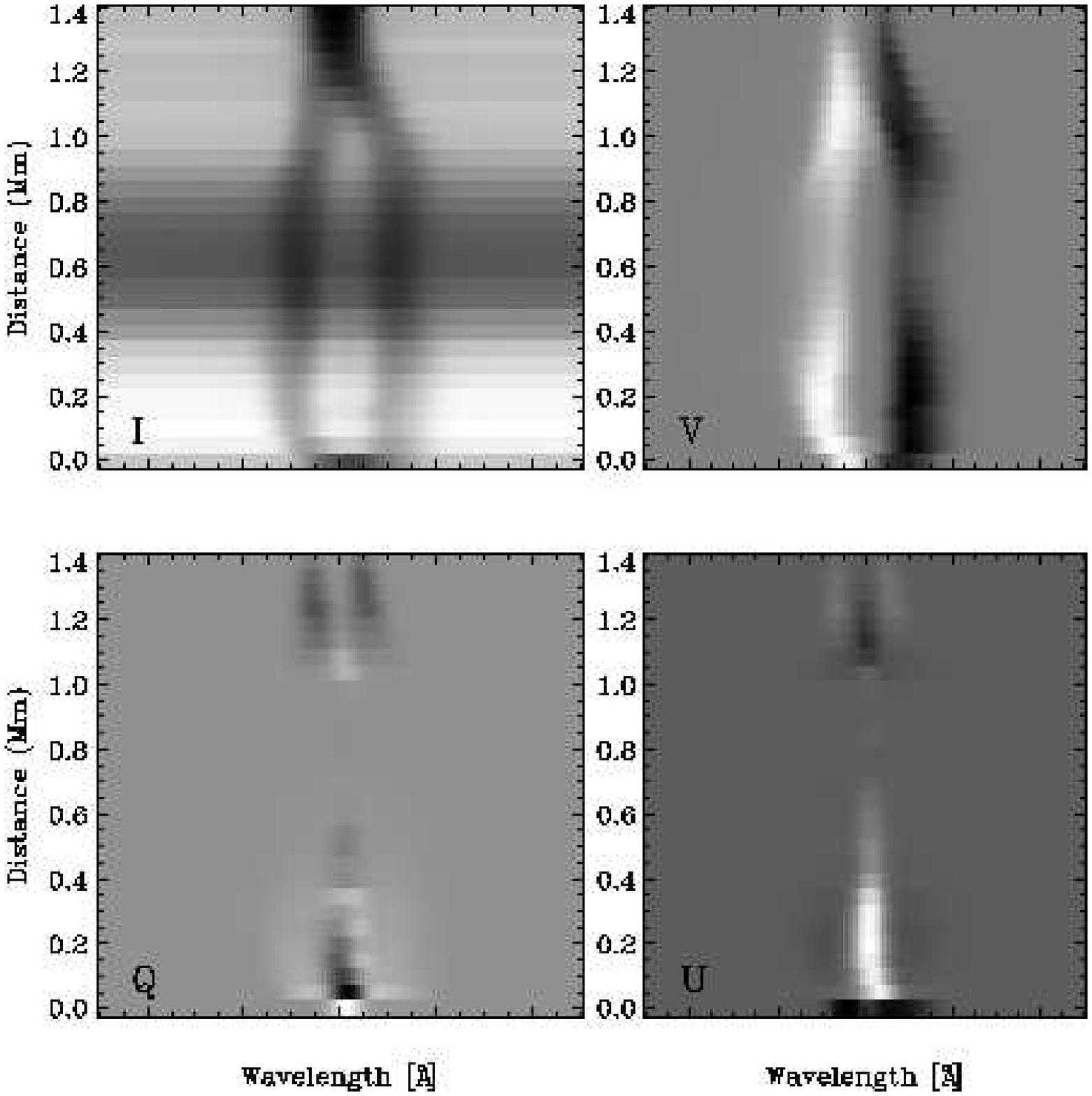,width=8.0cm}}
\caption{Stokes profiles of FeI 630.1 nm (right) along a slit across
a micropore, whose intensity image is shown on the left.}
\label{micropore}
\end{figure}

The surface magnetic field can also be determined by analyzing stokes
component line profiles.  An example of the stokes profiles along a
slit across a micropore in the simulation is shown in
Fig.~\ref{micropore}.

\subsection{P-Modes}

\begin{figure}[!htb]
\centerline{
\psfig{figure=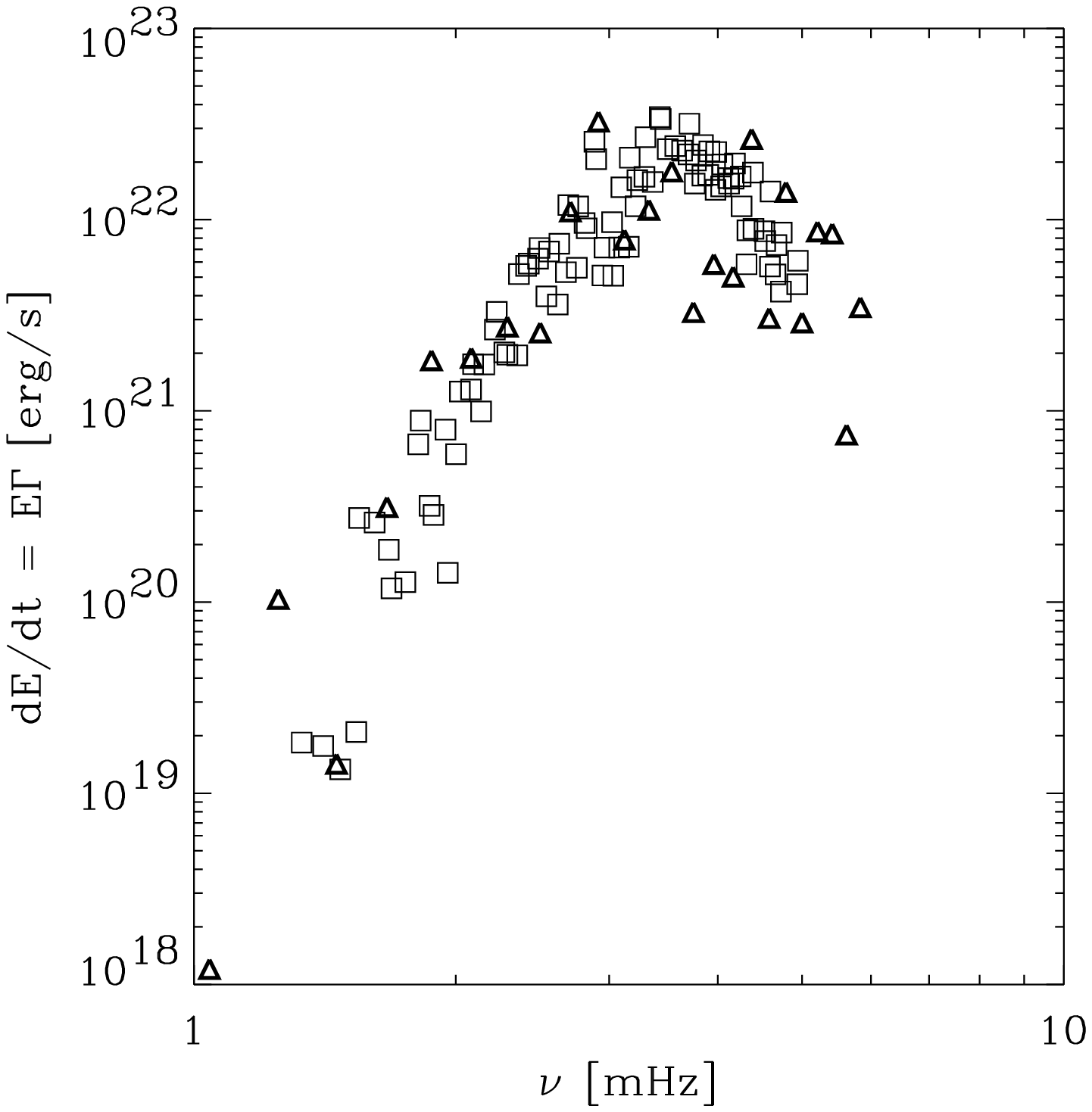,width=5.9cm}
\quad
\psfig{figure=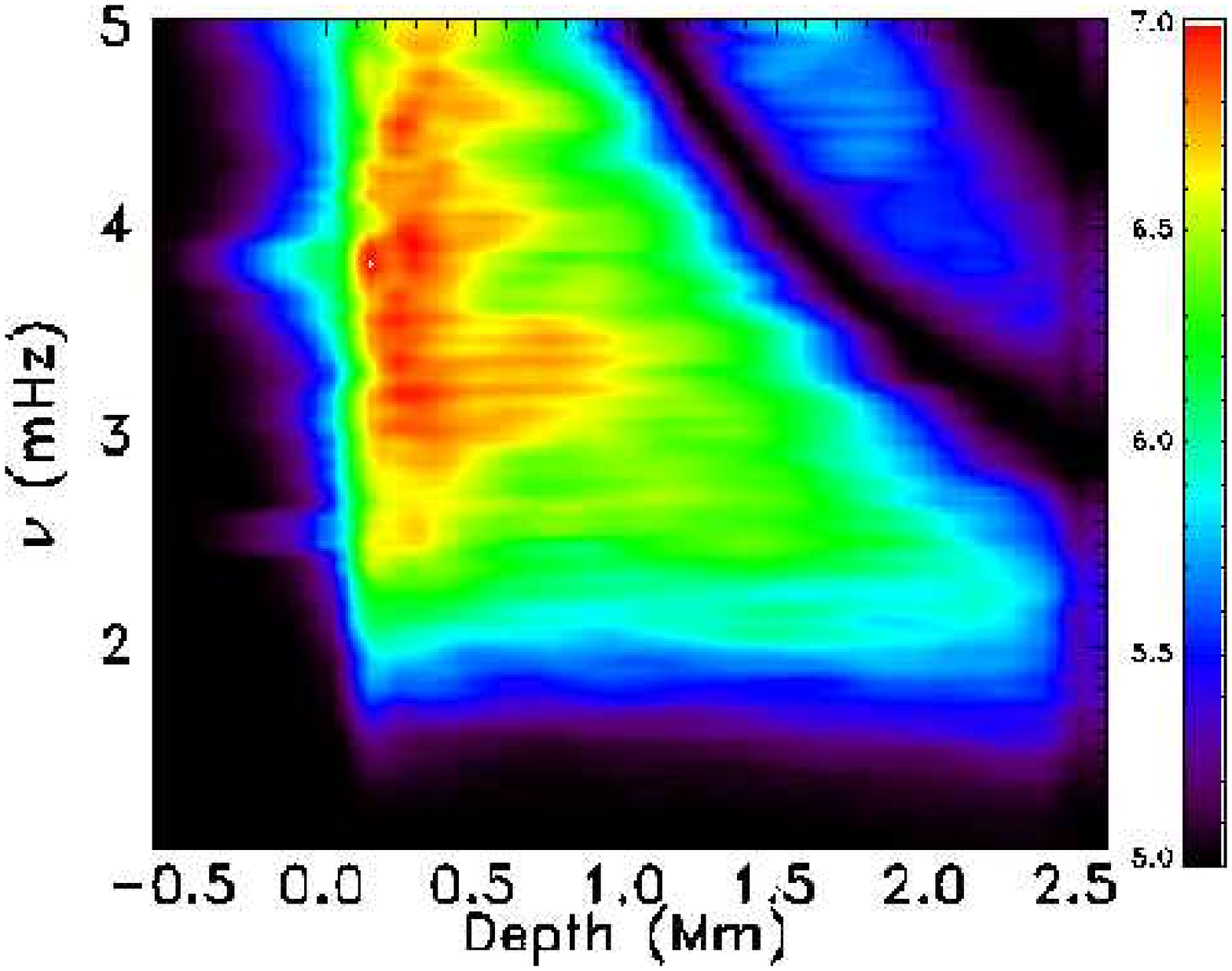,width=7.2cm}}
\caption{Solar p-modes are excited by the work of turbulent pressure
and non-adiabatic gas pressure fluctuations.  The excitation is confined
closer to the surface as the frequency increases.}
\label{mode-excitation}
\end{figure}

The solar p-modes are excited by the work of turbulent pressure and
non-adiabatic gas pressure (entropy) fluctuations near the top of the
convection zone (Fig~\ref{mode-excitation}).  The turbulent pressure is
due to the convective motions.  The entropy fluctuations result from
the instantaneous local imbalance between the convective heating and
radiative cooling.  P-mode excitation decreases at low frequencies
because of the increasing mode mass and decreasing mode compression as
the frequency decreases.  Excitation decreases at high frequencies
because the pressure fluctuations produced by the convection fall off
with increasing frequency.  The modes are excited fairly close to the
solar surface -- the higher the frequency the closer to the surface is
their excitation (Fig~\ref{mode-excitation})
(Stein \& Nordlund 2001).

\begin{figure}
\centerline{
\psfig{figure=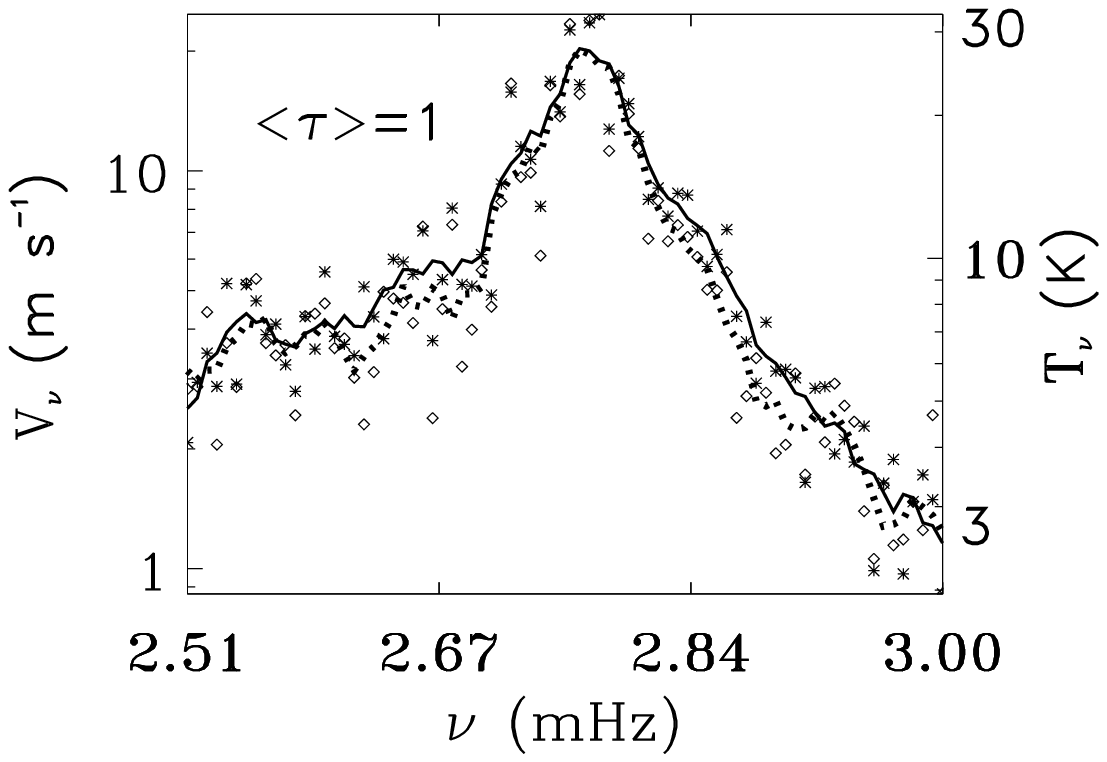,width=6.6cm}
\quad
\psfig{figure=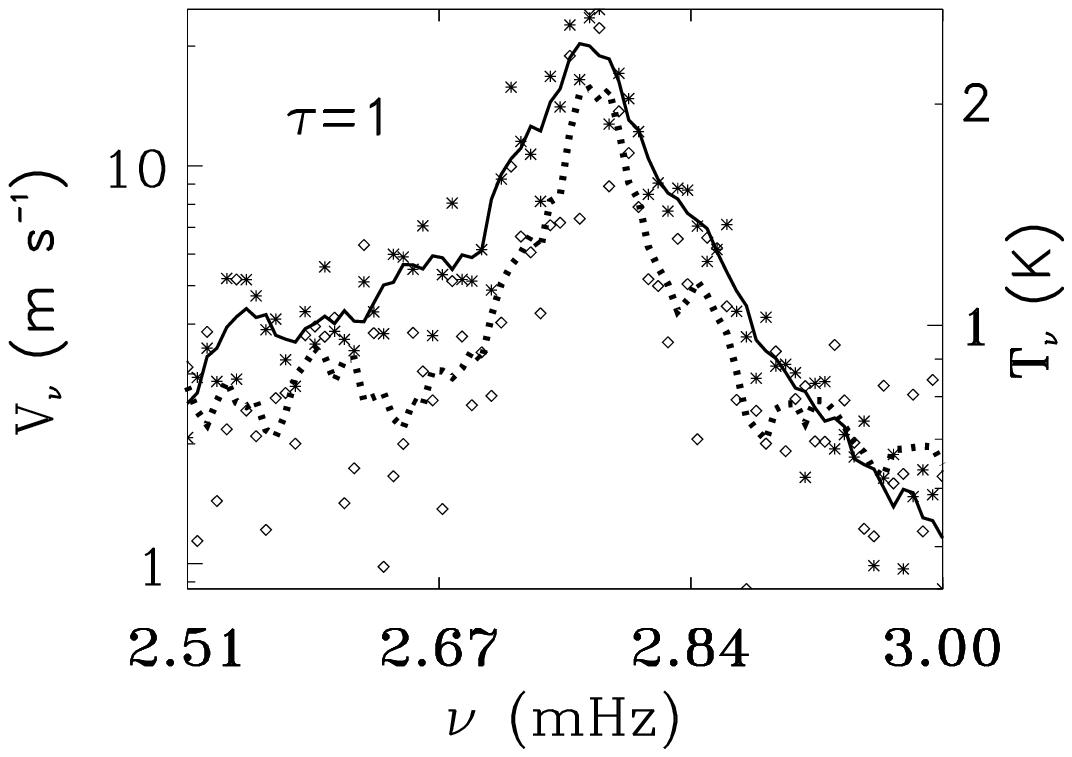,width=6.6cm}}
\caption{Velocity (solid) and temperature (dotted) spectra at fixed
geometrical depth $<\tau>=1$ (left) and optical depth $\tau=1$ (right),
for the first non-radial fundamental mode (corresponding to solar mode
with $\ell=740$).  Radiation transfer reduces the observed temperature
fluctuations by over an order of magnitude and reverses the asymmetry
of the temperature profile.
}
\label{V-Ttau}
\end{figure}

\begin{figure}
\centerline{
\psfig{figure=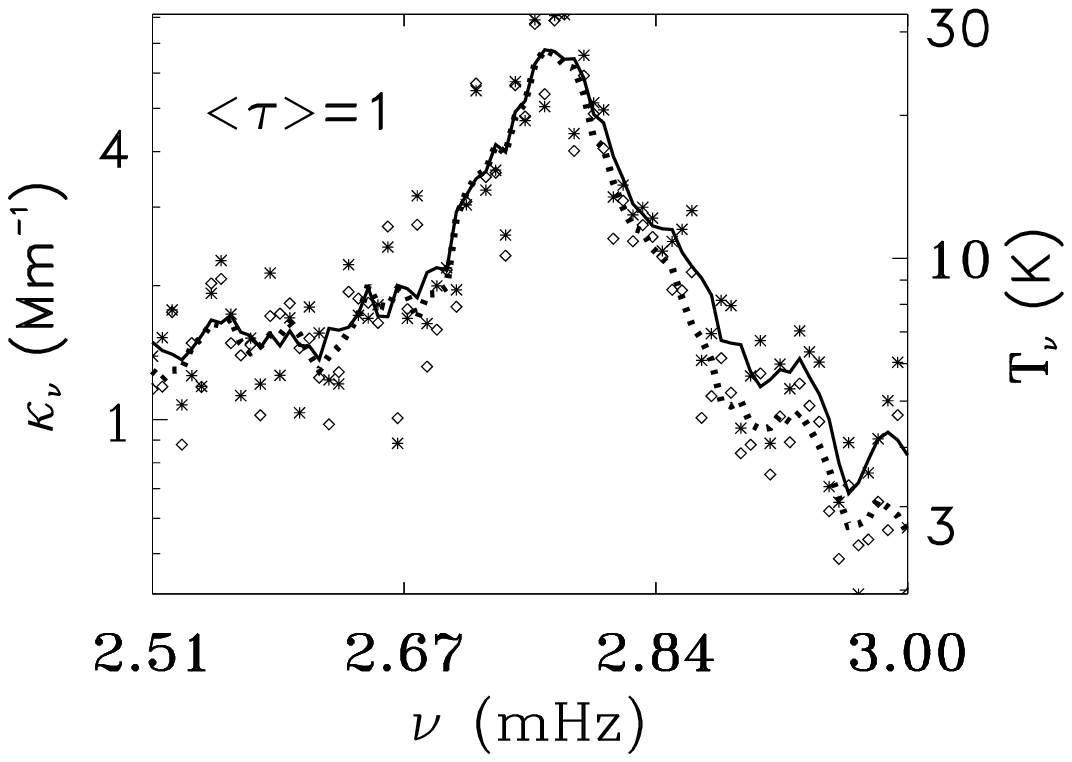,width=6.6cm}
\quad
\psfig{figure=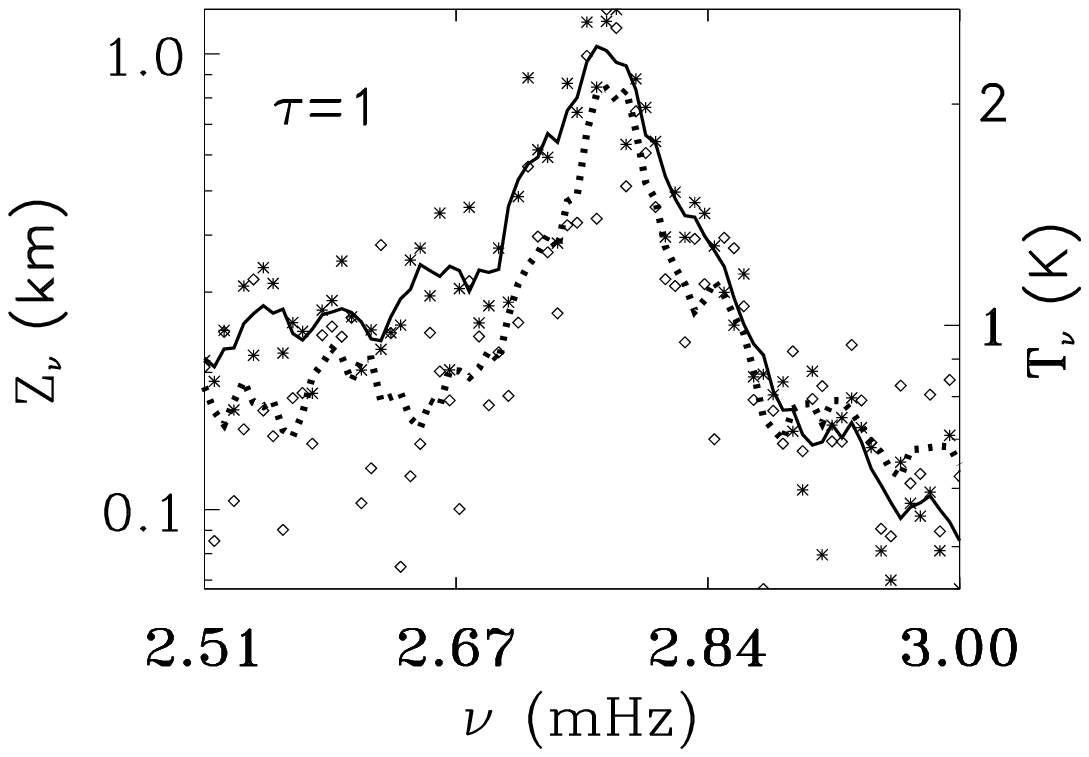,width=6.6cm}}
\caption{Opacity (left) and height of optical depth unity (right).
Larger temperature fluctuations (dotted) at frequencies below the mode
peak generate larger opacity fluctuations, which in turn produce larger
fluctuations in the height of the $\tau =1$ surface.  This decreases the
temperature fluctuations more below the peak than above it and reverses
the asymmetry of the intensity with respect to the velocity.
}
\label{K-Z-T}
\end{figure}

The velocity and intensity spectra of the p-modes have asymmetric peaks.
The velocity has more power on the low frequency side of the peak and
the intensity has more power on the high frequency side of the peak.
The velocity and intensity measured at a fixed geometrical depth have
the same asymmetry, which depends on the locations of the excitation
source and the observations (Kumar \& Basu 1999, Georgobiani et al. 2000).
The intensity asymmetry gets reversed from
that of the temperature by radiation effects.  The dominant H$^-$
opacity is very temperature sensitive.  The oscillation induced
temperature fluctuations are larger on the low frequency side of the
resonance and produce opacity changes that vary the location of
radiation emission ($\tau=1)$ in a way that reduces the magnitude of the
temperature fluctuations more on the low frequency side of the resonance,
and thus reverses the asymmetry of the observed temperature
fluctuations (Figs~\ref{V-Ttau}, \ref{K-Z-T}).

\section{Discussion}

In order to make realistic numerical simulations it is necessary to
include the relevant realistic physics.  In the case of the solar
atmosphere this means including an appropriately realistic treatment of
the radiative losses.  Many techniques are available to solve for
radiation transport.  The problem is to devise a method that is
efficient enough to be applied at each time step in a dynamic
calculation.

We have described a method developed by Nordlund that relies on
reducing the number of frequencies that need to be treated to a bare
minimum of 4, by using a multi-group binning of the wavelengths
according to their opacity.  In addition a minimal number of rays are
employed, but they are rotated to cover the computational domain.
Because we are only interested in the photosphere, we also can assume
LTE.  Some of the approximations we use, such as assuming that all the
opacities have the same depth dependence, can be easily improved.  It
was instituted in a time when computer memory was small and the table
size needed to be minimized.  It would be nice to replace the
approximation of multi-group opacities with an opacity sampling method,
so as to include the effect of Doppler shifts on the radiation
absorption, but the cost will be many more frequencies that need to be
solved.  For the Sun we are lucky that the radiation heating/cooling
time scale is of the same order as the dynamical time scale, so that an
explicit solution of the radiation transport is possible.  For giant
stars, for instance, where the dynamical time scale is much longer than
the Sun's, explicit solution of the radiation transport will reduce the
possible time step size.

For other investigations, other physics will be needed.  To include the
chromosphere it is necessary to include at least the dominant non-LTE
effects of scattering and hydrogen ionization.   This is the next
project we are working on in collaboration with Mats Carlsson, Viggo
Hansteen and Andrew McMurray of Oslo University.

\acknowledgments
The work of RFS was supported in part by NASA grant NAG 5 9563, NSF grants
AST 9819799 and ATM 9988111.  {\AA}N was supported in part by the Danish Research
Foundation, through its establishment of the Theoretical Astrophysics Center.
The calculations were performed at the National Center for Supercomputer
Applications, which is supported by the National Science Foundation, at
Michigan State University and at UNI$\bullet$C, Denmark.  This valuable
support is greatly appreciated.

\end{document}